\documentclass[aps,amsmath,amssymb,pre,twocolumn,showpacs,superscriptaddress]{revtex4}
\usepackage{graphicx}
\usepackage{lineno}
\usepackage{natbib}

\begin{document}
\title{Detrending Moving Average Algorithm: \\ Frequency Response and Scaling Performances }
\author{Anna Carbone}
\email{anna.carbone@polito.it}
\affiliation{Department of Applied Science and Technology, Politecnico di Torino, Corso Duca degli Abruzzi 24, 10129 Torino, Italy}
\author{Ken Kiyono}
\email{kiyono@bpe.es.osaka-u.ac.jp}
\affiliation{Graduate School of Engineering Science, Osaka University, 1-3 Machikaneyama-cho, Toyonaka, Osaka 560-8531, Japan}%

\date{\today}

\begin{abstract}
The Detrending Moving Average (DMA) algorithm has been widely used in its several variants for characterizing long-range correlations of random signals and sets (one-dimensional sequences or high-dimensional arrays) either over time or space. In this paper, mainly based on analytical arguments, the scaling performances of the centered DMA, including higher-order ones, are investigated by means of a continuous time approximation and a frequency response approach. Our results are also confirmed by numerical tests.
The study is carried out for higher-order DMA operating with moving average polynomials of different degree. In particular, detrending power degree, frequency response, asymptotic scaling, upper limit of the detectable scaling exponent and finite scale range behavior will be discussed.
\end{abstract}


\maketitle

\section{Introduction}

 To investigate whether the intensity of some relevant quantity is characterized by increasing or decreasing trends is a common goal to many research areas. In the simplest operational definition, trends are observed when a regression estimated over a data subset has a not negligible slope.
For example in climatology, based on regular measurements from weather stations and satellite
data, temperature trends are estimated
both locally and globally \cite{Zhang2012,Tamazian2015,Yuan2015,Bromwich2013}. In finance and economics, technical rules and visualization tools based on moving average trends are under continuous investigation and improvement \cite{Lo,Menkhoff,Longstaff,Daia}. A main issue in the application of trend estimates is related to the assumption of the model describing the underlying evolution process (e.g.~linear or exponential). Another critical aspect is the ability to distinguish whether the trend or other stochastic component embedded in a nonstationary time series arises from the intrinsic system dynamics or from external forcing drives (see e.g.~\cite{Tamazian2015}).
 Hence, the development of techniques (having the simultaneous ability of simulating trends and estimating long- and short-range  correlations of stochastic data sets) and the assessment of their performances is of relevance to diverse scientific communities.
\par
As a way to characterize nonstationary data with trend, the detrended fluctuation analysis (DFA) ~\cite{peng1994mosaic,peng1995quantification},
and the detrending moving average (DMA)
analysis ~\cite{alessio2002second,carbone2004analysis,arianos2007detrending,arianos2011self} have been proposed to quantify long-range autocorrelations,
 multifractal features ~\cite{GuPRE2010,GrechAPPB2005,BashanPhysA2008,MaPRE2010,Drozdz}, cross-correlation ~\cite{Podobnik,ArianosJstat2009}
 and higher dimensional fractals~\cite{GuPRE2006,CarbonePRE2007,TurkPRE2010,CarbonePRE2010} either in the time or in the space domain.
     According to the DFA, the time series is first divided in boxes of equal lengths, then trends are estimated as least-squares polynomial fitting of different orders $m$ in each non-overlapping and equally spaced box of length $n$.
  The DMA algorithm has been proposed as an alternative technique to quantify long-range correlations.
  In the frequency domain, the power spectral analysis is a well-established methodological framework~\cite{percival1993spectral,hamilton1994time}. By estimating the slope of the log-log plot of the power spectral density (PSD), a wide range of scaling behavior can be characterized. However, power spectral analysis may provide spurious estimates caused by the nonstationarity of time series, such as embedded trends and heterogeneous statistical properties.
  \par
  Statistical performance (e.g. effects of nonstationarity, nonlinear filters and extreme data loss) of the DFA and DMA have been investigated by a number of comparative studies
   ~\cite{alvarez2005detrending,chianca2005fourier,hu2001effect,chen2002effect,xu2005quantifying,chen2005effect}.
 However, the performances of DMA, especially higher order ones and the relation between DMA and power spectral analysis have not been investigated. 
 Therefore, the aim of this work is to understand such methodological features by an analytical approach based on the continuous time approximation and the single-frequency response of the DMA already adopted for DFA in~\cite{KiyonoPRE2015}.
 Based on the exact calculation of the single-frequency response function under the continuous time approximation,
the direct connection between DMA and power spectral analysis can be derived. Such relations are then exploited to derive a number of scaling performance features such as detrending power degree, frequency response, asymptotic behavior, upper limit of the scaling exponent and finite scale range behavior. The current work aims at providing clear mathematical reasoning for these properties and guiding principles to improve the detrending methodologies.
\par
The organization of this paper is as follows. In Section II, the main computational steps of the DMA are briefly recalled together with a brief discussion highlighting the relevance of the high-order detrending methods and the limits of detrending performance in scaling analysis. In Section III and in the Appendixes, the analytical derivation of the performance characteristic mainly based on frequency response reasoning is reported. A comparison of the results obtained for the DMA to the DFA ones is offered all through the manuscript.

\section{Detrending Moving Average algorithm}

 As already stated above, the focus in this work is on the DMA operating with moving average polynomials of different degree.  However, before entering the details of the present study, the basic elements of the DMA analysis will be summarized. The main ingredient of the DMA algorithm is the generalized variance $\sigma^2_{\rm DMA}(n)$ of the time series $\{y(i)\}_{i=1}^{N}$ with respect to the trend $\{\tilde{y}_n(i)\}$ at scale $n$:

\begin{equation}
\label{var}
\sigma^2_{\rm DMA}(n)  = \frac{1}{N-n+1} \sum_{i} \Big[y(i)-\tilde{y}_n(i)\Big]^2,
\end{equation}
where $\tilde{y}_n(i)$ is defined as a time-dependent average function of $y(i)$. In the simplest case, called backward DMA, $\tilde{y}_n(i)$ can be estimated as the ordinary  moving average:
\begin{equation}
\label{dmaapp1}
\tilde{y}_n(i)  =  \frac{1}{n}\sum_{k=0}^{n-1}y(i-k),
\end{equation}
and the range of the summation in Eq.~(\ref{var}) is from $n$ to $N$.
 For random walk-type processes with diffusive behavior, such as the fractional Brownian motion, the power-law increase of the root-mean square deviation $\sigma_{\rm DMA}(n)$ with the moving average window size $n$:
 \begin{equation}
\sigma_{\rm DMA}(n) \sim n^{\alpha},
\end{equation}
 \noindent
provides an estimate of the scaling exponent $\alpha$ and thus of the Hurst exponent $H$. For long-range correlated time series $\{x(i)\}_{i=1}^{N}$ with non-diffusive behavior, such as fractional Gaussian noise (fGn), the integrated series (cumulative sum), $y (i) = \sum_{j=1}^{i} x (j)$, as a sample path of a random walk driven by $\{x(i)\}$ is investigated and quantified in terms of  the scaling exponent $\alpha$.

\par
 As a generalization of  Eqs.~(\ref{var}) and (\ref{dmaapp1}), the high-order DMA (DMA$_m$)  has been proposed where the trends $\{\tilde{y}_n(i)\}$ are defined in terms of moving average polynomials of degree $m$ ~\cite{arianos2011self}. In this framework, the DMA algorithm with the conventional moving average given by Eq.~(\ref{dmaapp1}) is referred to as the zeroth order DMA (DMA$_0$).
 \par
\begin{figure}[bthp]
       \begin{center}
               \includegraphics[width = 0.8\linewidth]{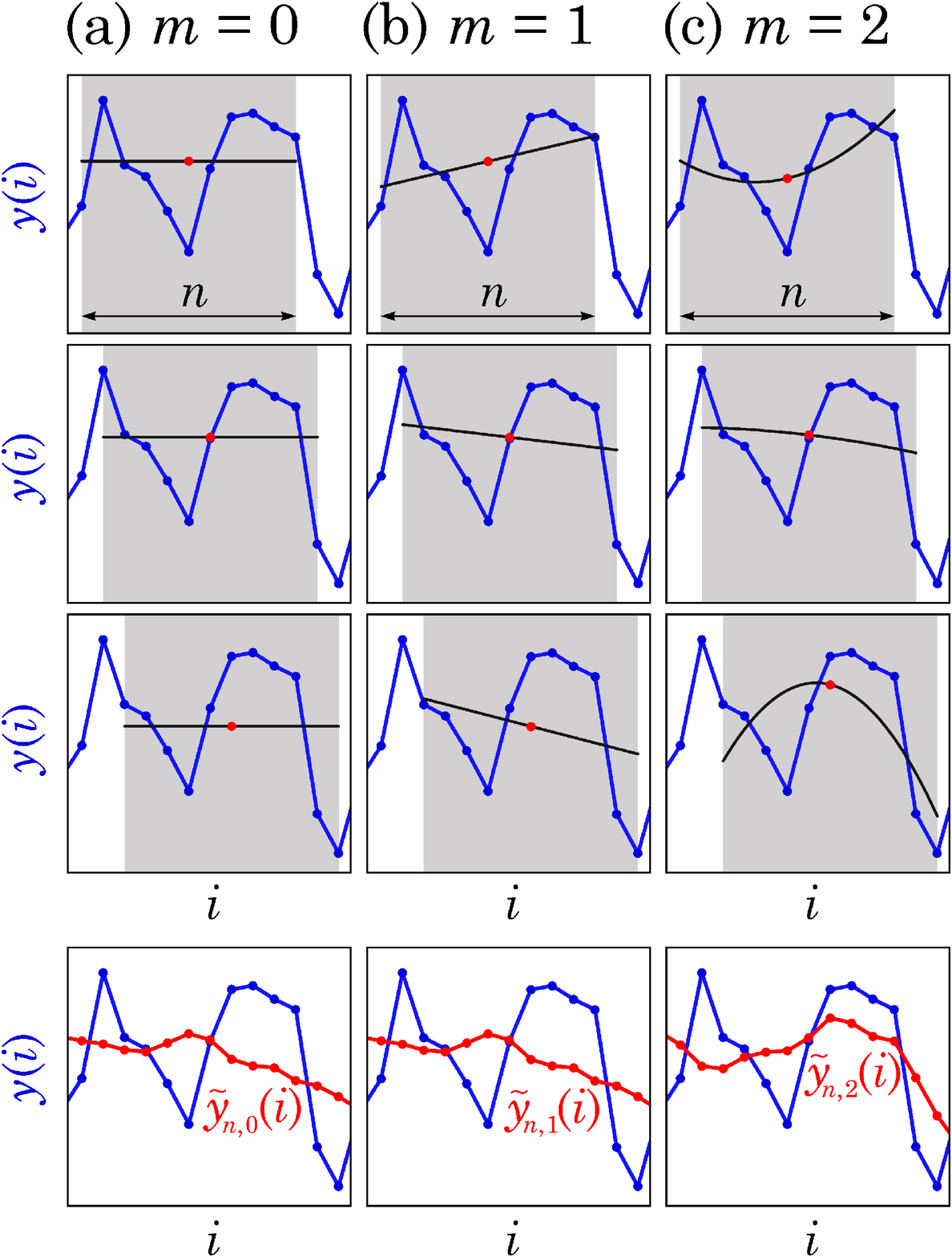}
               \caption{Illustration of the moving average polynomial of degree $m$ (bottom) for centered DMA$_m$ [Eqs.~(\ref{eq:a}) and (\ref{DMA_poly})], where $n$ is equal to 11 for all the cases. (a) Zeroth order ($m=0$, corresponding to the conventional moving average); (b) first order ($m=1$); and (c) second order ($m=2$). In top three rows, the center point (red points) on each polynomial fit (black lines) in a local moving average window (gray shaded areas) is defined as the higher-order moving average point, and the moving average polynomial $\{\tilde{y}_{n,m}(i)\}$ is calculated by point-by-point sliding of the window. On the other hand, for the backward DMA (given by Eq.~(\ref{dmaapp1})), the far right point in the window is defined as the moving average reference point. Note that, in the centered DMA, $\{\tilde{y}_{n,0}(i)\}$ in DMA$_0$ (bottom left) and $\{\tilde{y}_{n,1}(i)\}$ in DMA$_1$ (bottom middle) are identical.}
               \label{fig:ma}
       \end{center}
\end{figure}

 In the case of the centered DMA$_m$, the coefficients of the $m$th degree polynomials with moving average window over a range $[i - (n-1)/2, i + (n-1)/2]$, where  $n$ is assumed to be an odd number, are given by:
\begin{equation}
\left[
\begin{array}{c}
\tilde{a}_{n,0} (i) \\
\tilde{a}_{n,1} (i) \\
\vdots \\
\tilde{a}_{n,m} (i) \\
\end{array}
\right]= B_{m}^{-1}(i,n)\left[
\begin{array}{c}
\displaystyle \sum_{i'=i - (n-1)/2}^{i + (n-1)/2} y (i') \\
\displaystyle \sum_{i'=i - (n-1)/2}^{i + (n-1)/2} {i'} \, y (i') \\
\vdots \\
\displaystyle \sum_{i'=i - (n-1)/2}^{i + (n-1)/2} {(i')}^m y (i')
\end{array}
\right], \label{eq:a}\end{equation}
where $B_{m}^{-1}(i,n)$ is the inverse matrix of
\begin{eqnarray}
B_m (i,n) &=& \sum_{{i'}=i - (n-1)/2}^{i + (n-1)/2} \left[
\begin{array}{cccc}
1 & {i'} & \cdots &  ({i'})^m \\
 {i'} &  ({i'})^2 & \cdots &  ({i'})^{m+1} \\
\vdots & \vdots & \ddots & \vdots  \\
 ({i'})^m &  ({i'})^{m+1} & \cdots &  ({i'})^{2m} \\
\end{array}
\right]. \nonumber \\
\end{eqnarray}
\noindent
Thus, the moving average polynomial of degree $m$ is expressed as:
\begin{equation}
\tilde{y}_{n,m}(i) = \tilde{a}_{n, 0}\! \left(i \right) + \tilde{a}_{n, 1} \! \left(i \right) \, i + \cdots + \tilde{a}_{n, m} \! \left(i \right) \, i^m, \label{DMA_poly}
\end{equation}
where $i=1+(n-1)/2, \cdots, N-(n-1)/2$ and the coefficients $\left\{ \tilde{a}_{n, m} \right\}$ in Eq.~(\ref{DMA_poly}) are not constant, as they in fact are dependent on $i$ (see also Fig.~\ref{fig:ma}). To estimate $\sigma^2_{\rm DMA}(n)$ [Eq.~(\ref{var})] in the centered DMA$_m$, the moving average polynomial $\left\{\tilde{y}_{n,m}(i)\right\}$ is used as the trend $\left\{\tilde{y}_n(i)\right\}$ and the range of summation in Eq.~(\ref{var}) is from $1+(n-1)/2$ to $ N-(n-1)/2$. The current investigation is limited to even order centered DMA, infact the $2 k$th and $(2 k + 1)$th order centered DMA, where $k$ is a nonnegative integer, have been proven to be equivalent in~\cite{arianos2011self,PAPER2}.

\begin{figure}[htbp]
       \begin{center}
               \includegraphics[width = 1\linewidth]{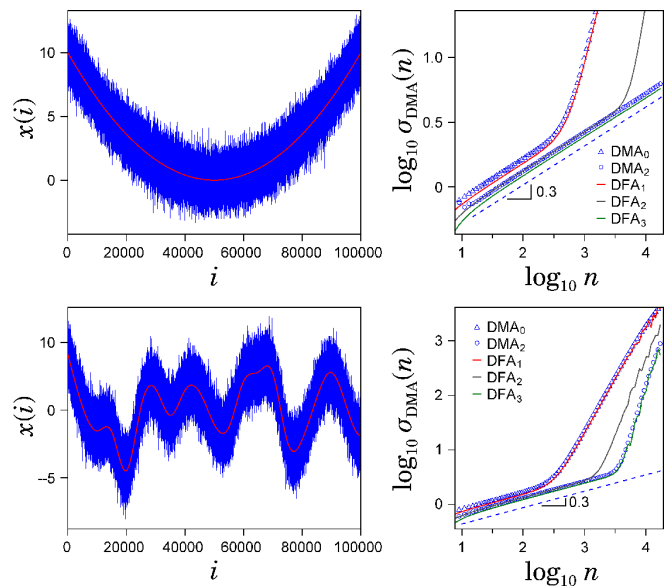}
               \caption{Illustration of trends estimated by the DFA$_3$ (left panels) and the centered DMA$_2$ (right panels). The box widths for the DFA$_3$  trend estimates are equal to 49 (top) and 99 (bottom). The moving average windows for the DMA$_2$ are equal to 49 (top) and 99 (bottom).}
               \label{fig:trend}
       \end{center}
\end{figure}

As an illustrative example in Fig.~\ref{fig:trend}, the trends calculated by using the third order DFA (left panels) and the second order DMA (right panels)  are shown. One can note that the DFA trend shows discontinuous jumps at the end points of each box. Conversely, the DMA trend exhibits seemingly continuous behavior. This is a crucial difference between DFA and DMA.

\par
To date, the performance of the zeroth order DMA algorithm (DMA$_0$) has been extensively studied  ~\cite{alvarez2005detrending,chianca2005fourier,hu2001effect,chen2002effect,xu2005quantifying,chen2005effect}. Conversely, the fundamental properties of the higher-order DMA (DMA$_m$ with $m\geq 1$) have not been completely investigated ~\cite{arianos2011self}. To fill this knowledge gap and understand the methodological performances of DMA$_m$, we study the properties of the high-order DMA based on analytical arguments and numerical tests. Furthermore, the results of the study will be thoroughly compared to DFA. In the remainder of this section, based on preliminary numerical results, we provide an overview of the issues involved in the performance of the high-order DMA.

\subsection{Importance of High-Order Detrending}\label{subsec:2A}

\begin{figure}[htbp]
       \begin{center}
               \includegraphics[width = 1\linewidth]{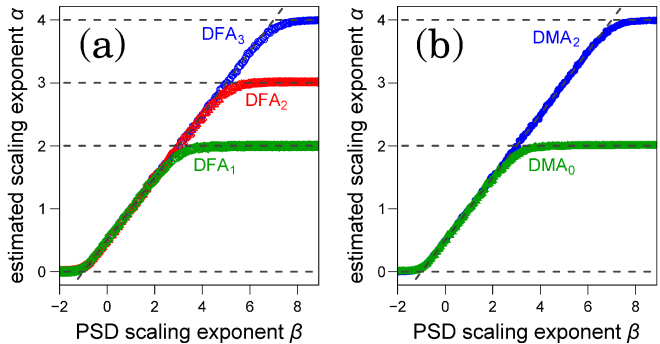}
               \caption{Scaling analysis of nonstationary time series with smooth trends. (Top) A sample time series generated by the sum of fractional Gaussian noise with $\alpha = 0.3$ and a quadratic trend component. (Bottom) A sample time series generated by the sum of fractional Gaussian noise with $\alpha = 0.3$ and a cubic spline trend in which 10 equally-spaced points with gaussian random numbers are interpolated by a cubic spline.
}
               \label{fig:nonstationary}
       \end{center}
\end{figure}

  When real-world long-range correlated series with nonstationary trends should be analyzed, high-order detrending is needed to detect the meaningful scaling exponents of the stochastic fluctuations embedded in the intrinsic or extrinsic trends. To illustrate this situation, let us consider the following case studies of artificially time series generated by:
  \begin{description}
    \item[(i)] the sum of a fGn with $H=0.3$ and a quadratic deterministic trend [Fig.~\ref{fig:nonstationary} (top left)]
    \item[(ii)]  the sum of a fGn with $H=0.3$  and a cubic stochastic trend [Fig.~\ref{fig:nonstationary} (bottom left)] obtained as the interpolation of  Gaussian generated points  by a cubic spline.
  \end{description}
In Fig.~\ref{fig:nonstationary} (top and bottom right), the log-log plots of the centered {\rm DMA}$_0$, {\rm DMA}$_2$ and {DFA}$_1$, {DFA}$_2$, {DFA}$_3$ 
vs $n$ are respectively shown.
\par
As one can note [Fig.~\ref{fig:nonstationary} (top right)], the plots of the {\rm DMA}$_0$ show spurious scaling behavior as suggested  by the slope much steeper than  $H=0.3$  at large values of $n$. The steep slope is due to the adverse effect of the deterministic quadratic trend component. Conversely, the {\rm DMA}$_2$ plot shows a scaling behavior with the correct slope $H=0.3$ over the whole range of $n$, which demonstrates that the  moving average  polynomial of second degree $\{\tilde{y}_{n,2}(i)\}$ is not affected by the deterministic quadratic trend. Similar results have been found for the higher-order DFA as shown in Fig.~\ref{fig:nonstationary} respectively for {DFA}$_1$, {DFA}$_2$ and {DFA}$_3$. In particular for the case (ii) [Fig.~\ref{fig:nonstationary}(bottom)], one can note that the upper boundary of the scaling range of {DFA}$_3$ (or {DMA}$_2$) is approximately one order larger than that of {DFA}$_1$ (or {DMA}$_0$).
As a conclusion, the general result of the investigation is that higher-order DFA and DMA can reduce the adverse effect of the quadratic and cubic trend, by extending the scaling range towards higher values of $n$ [Fig.~\ref{fig:nonstationary}].
\par
To date the detrending ability of the DMA with respect to embedded polynomial trends has not been analyzed and understood in depth. Therefore, in this paper, we will analytically show the detrending ability of DMA with respect to different trends and compare the performance with the DFA.
\par
As a final remark, given the relevance of the above issues, it is convenient to introduce a figure of merit for scaling methods to remove a polynomial trend that we will refer to as Detrending Power Degree.

\subsection{Upper limit of detectable scaling exponents}\label{subsec:2B}

Next, the relationship between the upper limit of detectable scaling exponent $\alpha$ and the degree of the moving average polynomial is investigated. Though one can expect that upper limit would exist \cite{KiyonoPRE2015}, this aspect has not been systematically studied and thus will be analytically derived by using the single-frequency response function of DMA. Before illustrating the analytical results, the numerical results obtained for both DMA and DFA will be briefly discussed here.
\par
\begin{figure}[htbp]
       \begin{center}
               \includegraphics[width = 1\linewidth]{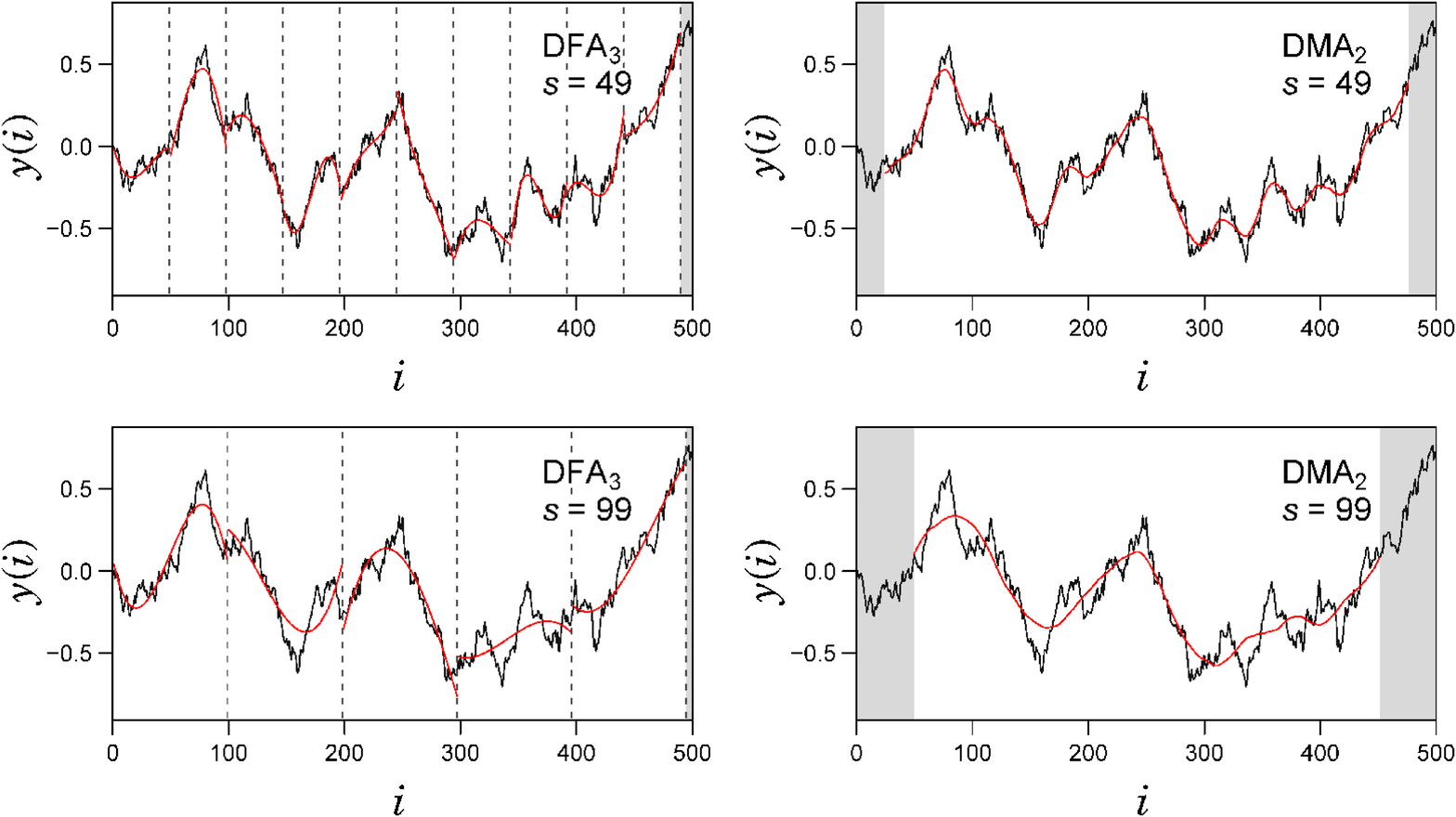}
               \caption{Scaling exponent $\alpha$ for a time series with  power spectral density $1/f^{\beta}$ analyzed respectively by (a)  first, second and third order DFA and (b) zeroth and second order DMA. Mean values of 30 samples are plotted. }
               \label{fig:limit}
       \end{center}
\end{figure}

Figure~\ref{fig:limit} (a) shows the scaling exponents $\alpha$ estimated by $m$th order DFA when sample time series with $1/f^{\beta}$ slope are analyzed. For the DFA, the relation $\alpha = (\beta + 1)/2$ holds in a range $0 < \alpha < m+1$, and the upper limit of detectable scaling exponent $\alpha$ is equal to $m+1$ \cite{KiyonoPRE2015}.
\par
On the other hand, as shown in Fig.~\ref{fig:limit} (b), the upper limit of detectable scaling exponents of DMA$_0$ and DMA$_2$ are respectively two and four.  Therefore, one can conclude that, both in DFA and DMA, higher-order detrending allow one to extend the upper limit of detectable scaling exponents.

\section{Analytical derivation of the DMA performance features}

In this section, the methodological performances, discussed and illustrated by using the numerical results briefly illustrate in the previous section, will be analytically derived. It is worth noting that by using the approach proposed in this section, it would be possible to study properties of a wide class of random walk analysis including the several variants of the DMA.

\subsection{Detrending Power Degree}\label{subsec:3A}

Here, we study the Detrending Power Degree of the centered DMA, whose relevance for scaling analysis was discussed in subsection \ref{subsec:2A}. In particular, we will show that the DMA order $m$ which is defined by the degree of the moving average polynomial $\{\tilde{y}_{n,m}(i)\}$, is also related to the degree of the detectable polynomial trend embedded in the long-range correlated time series. More precisely, it will be shown that, when the integrated time series $y(i)$ is analyzed, the $m$th order DMA can correspondingly remove polynomial trend with degree $m$ in the original time series $x(i)$.
\par
 Let us consider the sum of two uncorrelated time series $x_A(i)$ and $x_B(i)$, the superposition law of the mean square deviation holds \cite{Shao}:

\begin{equation}
\sigma^2_{\rm DMA} (n)_{A+B} = \sigma^2_{\rm DMA} (n)_{A} + \sigma^2_{\rm DMA} (n)_{B}, \label{eq:sp}
\end{equation}
\noindent
where $\sigma^2_{\rm DMA} (n)_{A}$, $\sigma^2_{\rm DMA} (n)_{B}$ and $\sigma^2_{\rm DMA} (n)_{A+B}$ denote the mean square deviation corresponding to $x_A(i)$, $x_B(i)$ and $x_A(i)+x_B(i)$, respectively. Therefore, if a time series is given by the sum of a fGn and a polynomial trend, the additive property of the mean square deviations holds. Therefore, the effect of the polynomial trend can be separated and investigated.
We consider the $m$th order centered DMA, where $m$ is assumed to be a nonnegative even integer. To simplify the calculation, we assume a continuous function of $t$ over the range $-T/2 \le t \le T/2$ with length (scale) $T$, and calculate the $m$th order moving average at $t = 0$. Note that, by parallel translation, arbitrary situation of the polynomial trend in a moving average window can be described in the following form. Thus, without loss of generality, by considering only a single point at $t=0$, we can study the general properties of DMA.
Let us consider a polynomial function with degree $q$, $x(t) = c_0 + c_1 t + \cdots + c_q t^{q}$, as the trend component in the original time series, and analyze its integrated function as:
\begin{equation}
y (t) = c_0 t + \frac{c_1}{2} t^2 + \cdots + \frac{c_q}{q+1} t^{q+1} = \sum_{k=0}^{q} \frac{c_k}{k+1} t^{k+1}.
\end{equation}
To calculate the value of the moving average polynomial at $t = 0$, one needs first to calculate the coefficients $\{a_k\}$ of the least-squares polynomial by minimizing
\begin{equation}
I\left( \{a_k\} \right) = \int_{-T/2}^{T/2} \! \left( \sum_{k=0}^{q} \frac{c_k}{k+1} t^{k+1} - \sum_{k=0}^{m} a_k t^{k} \right)^2 \, dt. \label{SqDevPoly1}
\end{equation}
Then, by using $\{a_k\}$ and substituting $t = 0$ into the integrand in Eq.~(\ref{SqDevPoly1}), the square deviation from the moving average polynomial at $t = 0$ is given by $a_0^2$. Finally, as shown in the Appendix \ref{derivation_a0}, we obtain:
\begin{equation}
a_0^2 = 0 \qquad {\rm for} \quad q \le m \hspace{5pt}, \label{eq:SqDevt0}
\end{equation}
which means that the moving average polynomial of degree $m$ coincides with $(q+1)$th degree polynomial trend $y(t)$ after integration. Thus, its generalized variance is equal to zero. In other words, if we evaluate the detrending power degree based on the order of the analyzed polynomial function $x(t)=\sum_{k=0}^{q} c_k\, t^q$, the detrending power degree of the centered DMA$_m$ is equal to $m$. In contrast, the detrending power degree of $m$th order DFA is equal to $m-1$.
\par
Analogously, for $q > m$, the square deviation from the polynomial trend takes a nonzero value. For instance, when $m=0$ and $q=1$, we get:
\begin{equation}
\sigma^2_{\rm DMA}(n) \approx \frac{c_1^2}{576} n^4,
\end{equation}
and, when $m=2$ and $q=3$, we get:
\begin{equation}
\sigma^2_{\rm DMA}(n) \approx \frac{9 c_3^2}{5017600} n^8.
\end{equation}
These results demonstrate that the polynomial trend with degree $q > m$ exhibits a spurious scaling behavior. Therefore, when the sum of a fractional Gaussian noise and a polynomial function with degree $q > m$ is analyzed by the centered DMA$_m$, a crossover in the plot of $\log \sigma_{\rm DMA}(n)$ vs $\log n$ appears according to the superposition law [Eq.~(\ref{eq:sp})].
\noindent
The numerical results plotted in Fig.~\ref{fig:nonstationary} confirm the above findings.

\subsection{Frequency Response}\label{subsec:3B}
Here, by using the power spectral density of a fGn and the frequency response of DMA, we will investigate the properties of DMA when fractional Gaussian noises are analyzed.
\par
It has been rigorously shown that the PSD of a fGn, increment process of a fractional Brownian motion with the Hurst exponent $H$, is given by  \cite{KOU2004,li2006rigorous,li2009fractal}:
\begin{equation}
S(\omega) = \sigma^2 \sin (H \pi) \Gamma (2 H + 1) |\omega|^{1-2H},  \label{psd:fGn}
\end{equation}
where $\omega$ is the angular frequency, $\sigma^2$ is the scale parameter of fGn, and $\Gamma$ is the gamma function.
\par
To evaluate the frequency response of DMA, we first calculate the single-frequency response function under a continuous time approximation. The single-frequency response function provides an analytical approximation of the generalized variance when a single-frequency signal component having amplitude $A$ and frequency $f$ is analyzed. Note that, as given in Appendix \ref{Appendix:EC}, it is possible to calculate the exact form of the single-frequency response function without the continuous time approximation. However, to simplify the calculation and to compare with the previous study of DFA \cite{KiyonoPRE2015}, we here use the continuous time approximation.
\par
 Let us thus consider the single frequency component, $x(t) = A \cos \left(2 \pi f t + \theta \right)$,  a continuous-time signal that after integration can be written as:
\begin{equation}
y(t) = \frac{A}{2 \pi f} \sin \left(2 \pi f t + \theta \right) \hspace{5pt},  \label{yt}
\end{equation}
where the constant integration term has been neglected.
For the purpose of simplicity, the following calculation will be limited to the interval $\left[- T/2, T/2 \right]$ of length $T$ corresponding to the scale $n$ in the discrete-time notation. To investigate the frequency response of the centered DMA, we estimate the square deviation at $t = 0$ with respect to the moving average polynomial of degree $m$. To gain more insight into the DMA, it is valuable to note the difference in the calculation of the single-frequency response between DFA and DMA. In the previous study on DFA \cite{KiyonoPRE2015}, to calculate the single-frequency response function, the mean square deviation in a partitioned window (box) over $\left[- T/2, T/2 \right]$ is assumed (see Eq.~(12) in \cite{KiyonoPRE2015}), because the statistical property of each window is identical. In contrast, in the case of DMA, the square deviation at only a single point ($t=0$) is assumed, because the statistical property at each point is identical. This may be an advantage of DMA, because the estimate of the mean square deviation in DMA has a better statistical symmetry than that in DFA.

In the centered DMA, the moving average polynomial is obtained by minimizing the following function:
\begin{eqnarray}
I(\{a_0, a_1, \cdots, a_m \}) &=& \int_{- T/2}^{T/2} \! \left\{\frac{A}{2 \pi f} \sin \left(2 \pi f t + \theta \right) \right. \nonumber \\
&& \left. - \sum_{i=0}^m a_i \, t^i \right\}^2 \, dt \hspace{5pt}.
\end{eqnarray}
where the coefficients $\{a_i \}$ of the polynomials are determined by solving the following equations:
\begin{equation}
\frac{\partial I(\{a_i \})}{\partial a_j} = 0,
\end{equation}
with $i, j = 0,1,\cdots, m$.
\par
\begin{figure}[htbp]
       \begin{center}
               \includegraphics[width = 1\linewidth]{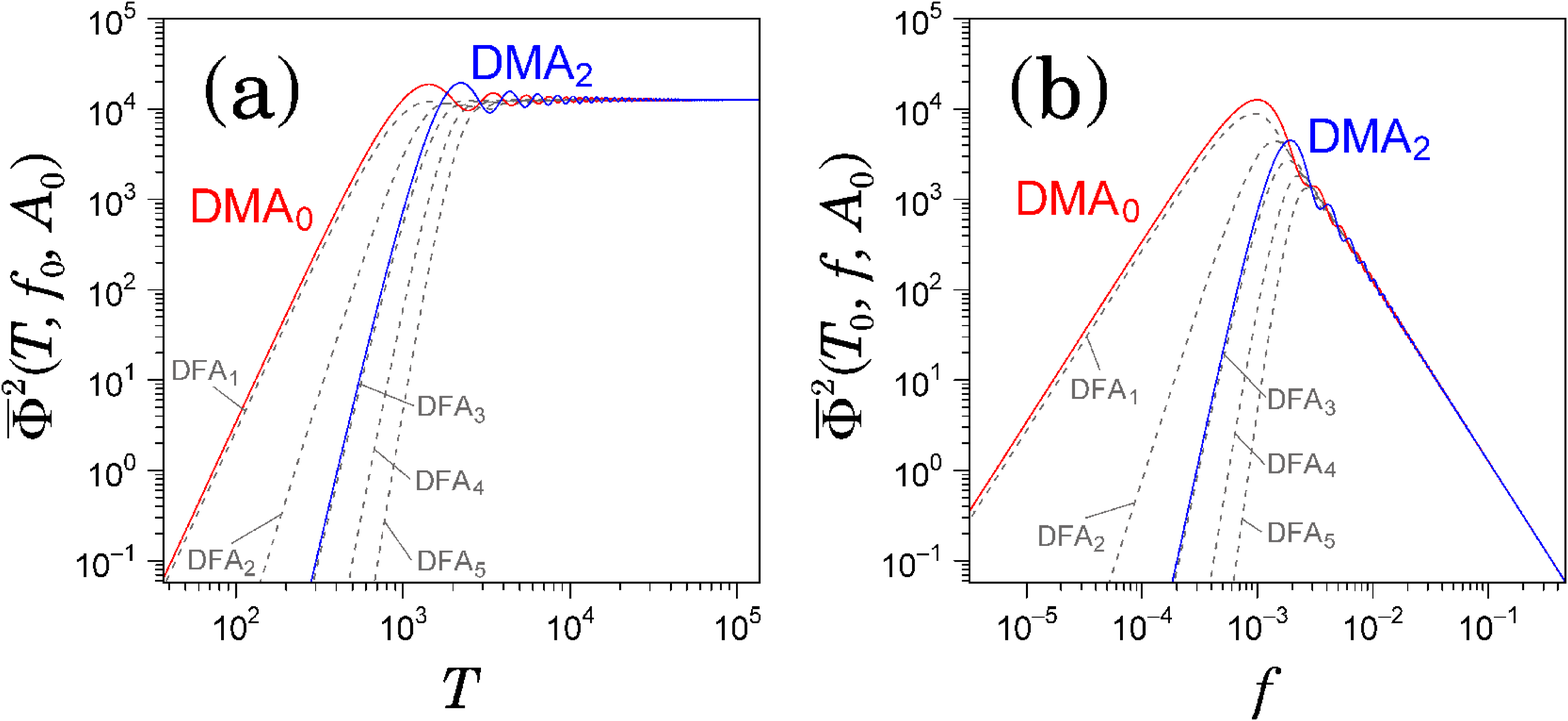}
               \caption{Single-frequency response functions $\overline{\Phi}^2 (T, f, A)$ of zeroth and second order DMA, where $T$ is the window length, $A$ is the amplitude of a Fourier component with frequency $f$. (a) $\overline{\Phi}^2 (T)$ versus scale $T$ with $f_0 = 10^{-3}$ and $A_0=1$. (c) $\overline{\Phi}^2 (f)$ versus frequency $f$ with $T = 10^3$ and $A_0=1$.  }
               \label{fig:freq_res}
       \end{center}
\end{figure}

The square deviation $\Phi^2$ with respect to the moving average polynomial at $t = 0$ is given by
\begin{equation}
\Phi^2_{t=0} (T, f, A, \theta) = \left( \frac{A}{2 \pi f} \sin \theta - a_0 \right)^2 \hspace{5pt}, \label{eq:Phi2}
\end{equation}
that, by averaging the phase $\theta$ in over $[0, 2 \pi]$, could be approximated by:
\begin{equation}
\overline{\Phi}^2 (T, f, A) = \frac{1}{2 \pi} \int_{0}^{2 \pi} \Phi^2_{t=0} (T, f, A, \theta)\, d \theta, \label{eq:Phi2ave}
\end{equation}
where we refer to $\overline{\Phi}^2$ as the single-frequency response function (more analytical details about $\overline{\Phi}^2$ are shown in the Appendix C).
The square root of $\overline{\Phi}^2 (T=n, f, A)$ can provide the analytical approximation of the $\sigma_{\rm DMA} (n)$ when a single-frequency component is  analyzed. From the curves plotted in Fig.~\ref{fig:freq_res}, one can note that the DMA$_0$ shows a single-frequency response similar to the DFA$_1$ and that the DMA$_2$ exhibits a single-frequency response similar to the DFA$_3$.  The fine structure of the single-frequency response is informative of the fundamental properties and the performance of the DMA  in the time and frequency domain.

\subsection{Asymptotic Scaling Behavior}\label{subsec:3C}

Here, we will demonstrate that, on the basis of the single-frequency response calculated in subsection \ref{subsec:3B}, the PSD of a linear stochastic process can be converted into the root mean square deviation $\sigma_{\rm DMA} (n)$.
\par
If the amplitude spectrum $\{A (f_k)\}$, where $f_k$ is the frequency of $k$th harmonic component, is known, the single-frequency response function $\overline{\Phi}^2$  provides an estimate of $\sigma_{\rm DMA} (n)$ as follows:
\begin{equation}
\sigma_{\rm DMA} (n) = \left[\sum_{k=1}^{\lfloor N/2 \rfloor} \! c(f_k) \, \overline{\Phi}^2 (n, f_k, A(f_k)) \right]^{1/2} \hspace{5pt}, \label{PSD2F}
\end{equation}
where $c(f_k)$ describes the effect of discrete time sampling:
\begin{equation}
\label{eq:c}
c(f) = \frac{(\pi f)^2}{\sin^2 (\pi f)} \hspace{5pt}.
\end{equation}
(see Ref.~\cite{KiyonoPRE2015} for a detailed derivation of Eqs.~(\ref{PSD2F}) and (\ref{eq:c})).
\par
Furthermore, by using Eqs.~(\ref{psd:fGn}) and (\ref{PSD2F}), the scale dependence of $\sigma_{\rm DMA} (n)$ for fGn can be analytically estimated.
As a representative example, we consider the asymptotic behavior of the  DMA$_2$.
When $n \ll 1/f$, the single-frequency response function $\overline{\Phi}^2$ [Eq.~(\ref{eq:Phi2ave})] of DMA$_2$ can be expanded as:
\begin{equation}
\overline{\Phi}^2 (n, f, A) = {C_{1} A^2 f^6 n^8} + O \left(n^{10} \right), \label{eq:phi_small}
\end{equation}
with the constant $C_{1} = \pi^6/627200$.
On the other hand, by taking the limit $n \to \infty$, we obtain
\begin{equation}
\lim_{n \to \infty} \overline{\Phi}^2 (n, f, A) = \frac{C_{2}A^2}{f^2 }. \label{eq:phi_inf}
\end{equation}
with the constant $C_{2} = (8 \pi^2)^{-1}$.
\noindent
Based on Eqs.~(\ref{eq:phi_small}) and (\ref{eq:phi_inf}), $\overline{\Phi}^2 (n, f, A)$ can be separated in two branches by:
\begin{equation}
\overline{\Phi}^2 (n, f, A) \approx  \left\{
\begin{array}{ll}
\displaystyle {C_{1} A^2 f^6 n^8}& \ {\rm for \ } n < \displaystyle  \frac{f_c}{f} \\
\\
\displaystyle \frac{C_{2}A^2}{f^2 } & \ {\rm for \ } n \ge \displaystyle  \frac{f_c}{f} \
\end{array}
\right. , \label{eq:phi_approx}
\end{equation}
where $f_c = \frac{280^{1/4}}{\pi}$.
If the power spectral density of a discrete sample of fGn is  a $1/f$-sloped function:
\begin{equation}
S_x (f) = \frac{A_0^2}{f^{\beta}}
\end{equation}
for $f \le 1/2$,  $\sigma_{\rm DMA} (n)$  can be estimated by assuming Eq.~(\ref{eq:phi_approx}) and $c(f) \approx 1$ as
\begin{widetext}
\begin{eqnarray}
\sigma_{\rm DMA}^2(n) &\sim& \int_{0}^{1/2} \! \left. \overline{\Phi}^2 \left(n, f, A \right) \right|_{A^2 = A_0^2 / f^{\beta}} \, df \nonumber \\
&\approx& C_{1} \int_{0}^{f_c/n} \! f^{6} n^{8} \left( A_0^2\, f^{-\beta}\right) \, df + C_{2} \int_{f_c/n}^{1/2} \! \frac{1}{f^{2}} A_0^2 f^{-\beta} \, df \nonumber \\
&=& \left( C_{1} \frac{f_c^{7-\beta}}{7 - \beta}+C_{2} \frac{f_c^{-\beta-1}}{\beta+1} \right) n^{\beta+1} - C_{2} \frac{2^{\beta+1}}{\beta + 1} \label{eq:F2calc}
\end{eqnarray}
\end{widetext}
where it is assumed $\beta < 7$. For $n \gg 1$ and $-1 < \beta < 7$, the first term of Eq.~(\ref{eq:F2calc}) is dominant, thus it results:
\begin{equation}
\sigma_{\rm DMA}^2(n) \sim n^{\beta+1}.
\end{equation}
which implies the following relationship:
\begin{equation}
\alpha = \frac{\beta+1}{2} \hspace{5pt}. \label{eq:scaling_ab}
\end{equation}

  Moreover since $\beta = 2H-1$ for fGn, it turns out that  $\alpha$ coincides asymptotically with $H$ of fGn. By using the same approach, analogous asymptotic laws can be derived for the high-order DMA$_m$.

\subsection{Upper limit of detectable scaling exponent}

The power-law tail structure of $\overline{\Phi}^2 (n, f, A) \sim f^{\gamma-2}\, n^{\gamma}$ for $n \ll 1/f$ determines the upper limit of the detectable scaling exponent $\alpha$ given by $\alpha < \gamma/2$. If $\alpha > \gamma/2$, namely $\gamma-\beta -1 <0$,  the $\sigma_{\rm DMA} (n)$ can be evaluated as:
\begin{eqnarray}
\sigma^2_{\rm DMA} (n) &\sim& \int_{0}^{f_c/n} \! f^{\gamma-2} n^{\gamma} \left( A_0^2\, f^{-\beta}\right) \, df \nonumber \\
&=& \left( A_0^2 \left[ \frac{f^{\gamma-\beta - 1}}{\gamma-\beta - 1} \right]_{f = 0}^{f_c/n} \right) n^{\gamma} \nonumber \\
&\approx& \left( A_0^2 \left[ \frac{f^{\gamma-\beta - 1}}{\gamma-\beta - 1} \right]_{f = 0}^{\epsilon } \right) n^{\gamma} \nonumber \\
&\sim& n^{\gamma}, \label{eq:F2upper}
\end{eqnarray}
where $0 < \epsilon \ll f_c/n$ is chosen such that:
\[
\left[ \frac{f^{\gamma-\beta - 1}}{\gamma-\beta - 1} \right]_{f = \epsilon}^{f_c/n} \ll \left[ \frac{f^{\gamma-\beta - 1}}{\gamma-\beta - 1} \right]_{f = 0}^{\epsilon }.
\]
Thus, the estimated scaling exponent is independent of $\beta$, and given by $\alpha = \gamma/2$.

 For $n \ll 1/f$, the single-frequency response of the DMA$_0$, DMA$_2$ and DMA$_4$ can be expanded in power of $n$ respectively as $\overline{\Phi}^2 (n, f, A) \sim f^2 n^4$, $\sim f^6 n^8$, $\sim f^{10} n^{12}$, respectively (see Appendix C).
Therefore, the scaling exponents $\alpha$ detectable by these methods are bounded by $2$, $4$ and $6$, respectively. Finally, one can conjecture that the upper limit of the detectable scaling exponent $\alpha$ by $2k$th order centered DMA, where $k$ is a nonnegative integer, would be $2k+2$.

\subsection{Finite scale range behavior}

The asymptotic behavior of the DMA when sample series of fGn are analyzed has been discussed on the basis of the single-frequency response function. In many practical situations, it is also important to understand the finite-range scale-dependence of $\sigma_{\rm DMA}(n)$. This issue will thus be investigated in this subsection.
\par
\begin{figure}[htbp]
       \begin{center}
               \includegraphics[width = 1\linewidth]{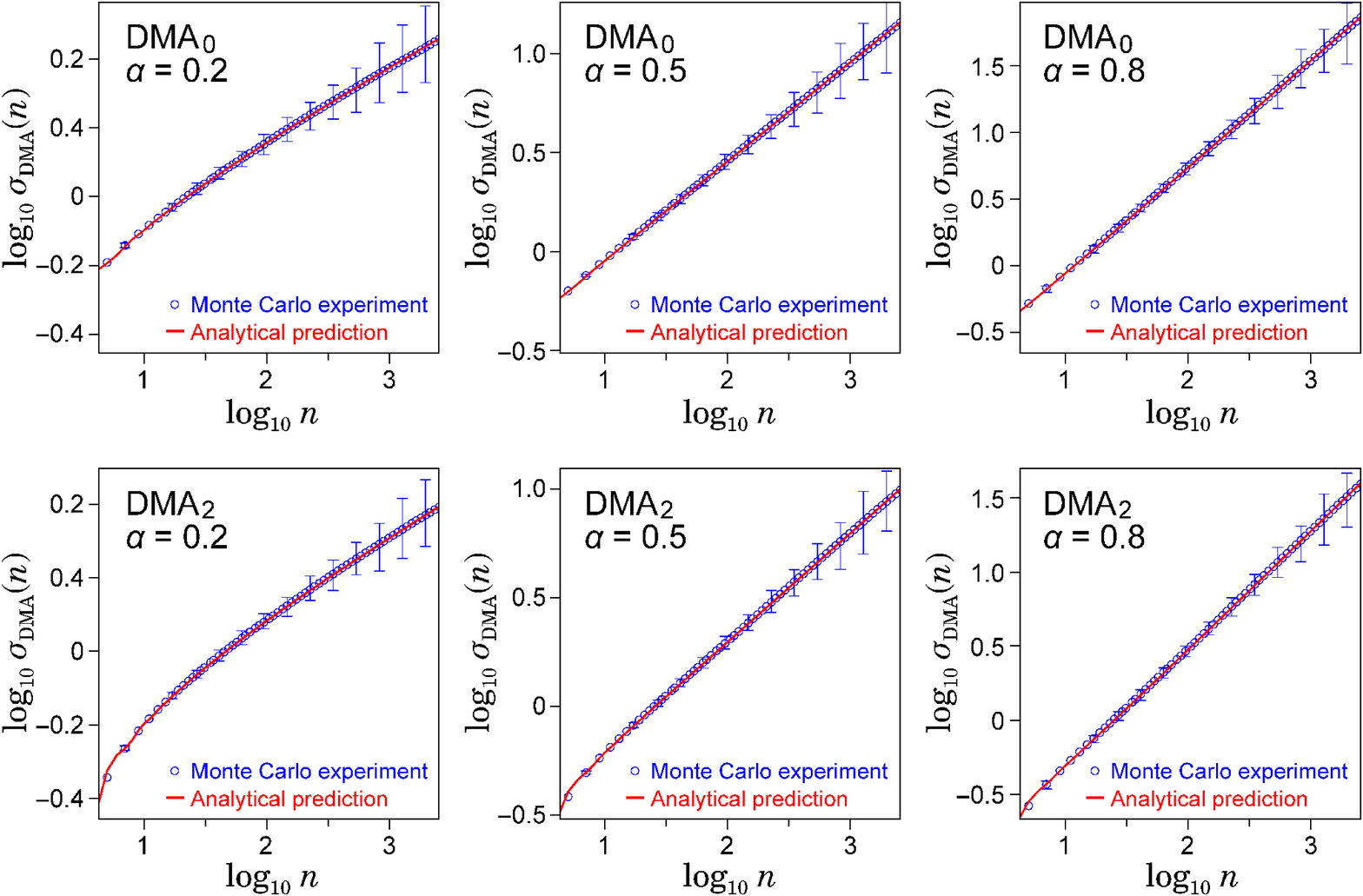}
               \caption{Scale dependence of the root mean square deviation from trend, $\sigma_{\rm DMA}(n)$ with order of the moving average polynomials equal to zero and two, obtained by Monte Calro (circles) and analytical (solid lines) approach. Artificially generated time series with given scaling exponent $\alpha$ are used for the numerical tests. }
               \label{fig:DMA_scale_dep}
       \end{center}
\end{figure}
\begin{figure*}[htbp]
       \begin{center}
               \includegraphics[width = 0.8\linewidth]{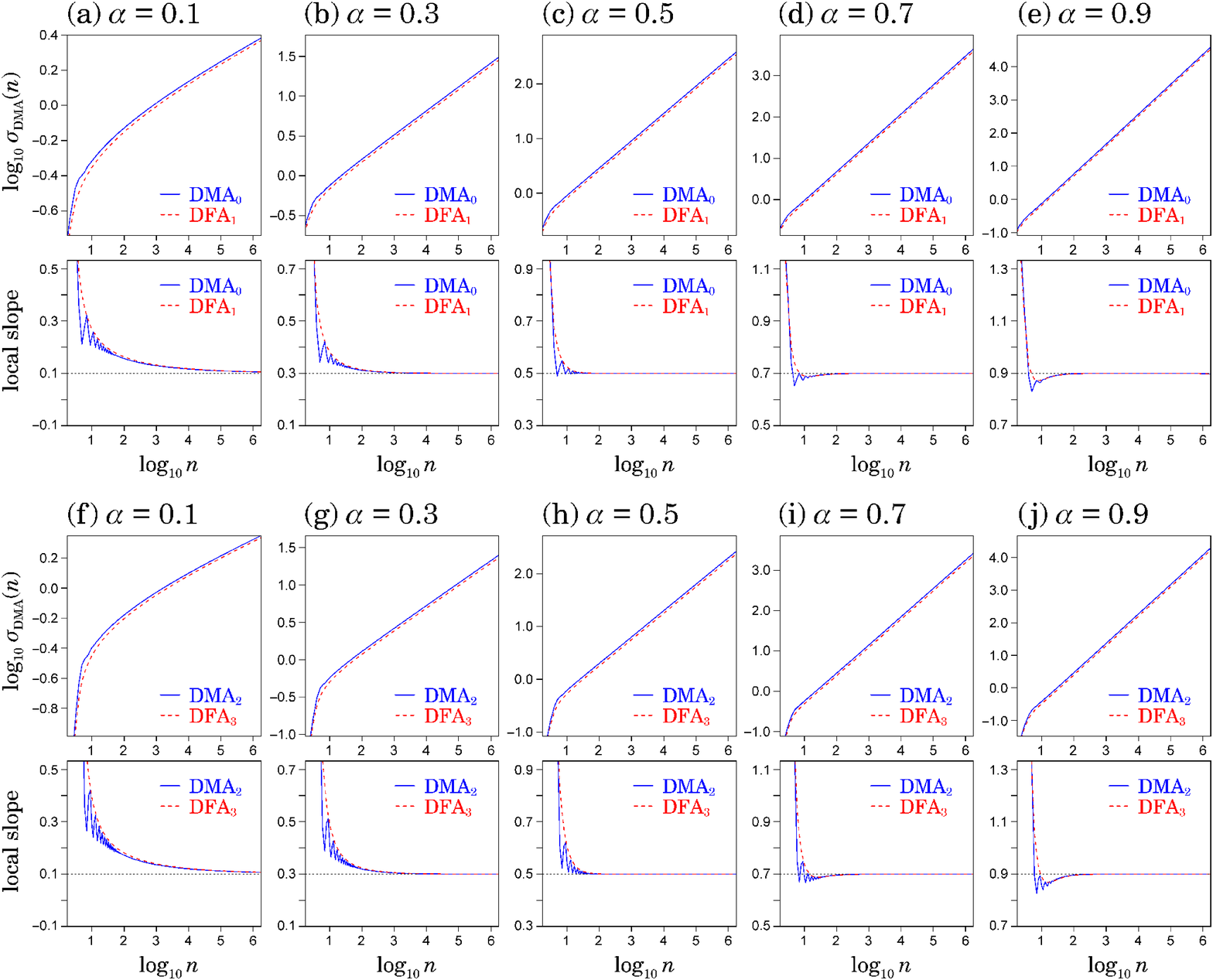}
               \caption{Scale dependence of $\sigma_{\rm DMA} (n)$ and its local slope. $\sigma_{\rm DMA} (n)$ is analytically evaluated based on the single-frequency response function. }
               \label{fig:finite}
       \end{center}
\end{figure*}

 By using Eqs.~(\ref{psd:fGn}) and (\ref{PSD2F}),  the scale dependence of $\sigma_{\rm DMA} (n)$  can be calculated for fGn. As shown in Fig.~\ref{fig:DMA_scale_dep} (solid lines), the predictions based on the analytical arguments are in good agreement with the numerical estimates obtained by the Monte Carlo approach. Through the analytical approach based on the single-frequency response function, we can precisely characterize the scaling behavior of fGn when analyzed by DMA. In Fig.~\ref{fig:finite}, the scale dependence of $\sigma_{\rm DMA} (n)$ and its local slope are shown together with results obtained for DFA.
The asymptotic convergence of the slope to the true value of $\alpha$ is very slow when $\alpha \approx 0$, as shown in Fig.~\ref{fig:finite}(a).   The local slopes at values of the scales $\log_{10} n < 1.5$ show oscillating behavior for DMA, which results from the oscillation seen in  the single-frequency response  of DMA [Fig.~\ref{fig:freq_res}(a)].  It is worthy of note that it would be very difficult to observe such fine structure of the $\sigma_{\rm DMA} (n)$ by using the Monte Carlo-based approach.
\par
 By the similarity of the single-frequency response functions of centered DMA$_m$ and DFA$_{m+1}$ (see Fig.~(\ref{fig:freq_res})), both methods show the similar finite scale range behavior in scales $\log_{10} n > 1.5$.

\section{Conclusion}

We studied methodological properties of high order centered DMA and on the basis of analytical arguments and numerical tests, we have demonstrated the following facts:

\begin{itemize}
                  \item $m$th order centered DMA can remove up to $m$th degree polynomial trend in the original time series before integration.
                  \item The single-frequency response functions of DMA$_0$ and DMA$_2$ have similar structure of DFA$_1$ and DFA$_3$, respectively.
                  \item The scaling exponent $\alpha$ estimated by centered DMA coincides asymptotically with the Hurst exponent $H$
                  \item The upper limit of the detectable scaling exponent $\alpha$ by $m$th order centered DMA is $m+2$
                \end{itemize}

It has been shown in \cite{alvarez2005detrending} that the detrending procedure in DFA is based on discontinuous polynomial fitting, which involves a nonlinear high-pass filter. Because of this nonlinearity, well established linear analysis methods, such as the frequency response based on frequency domain analysis, cannot be used on the DFA to investigate its methodological properties. To overcome this difficulty, a method based on the single-frequency response was recently proposed \cite{KiyonoPRE2015}. According to this method, the single-frequency response function is calculated through the analysis of a single-frequency component in the time domain, and can help to gain deeper insight into the performance of DFA including higher order cases \cite{KiyonoPRE2015}. In this paper, this approach has been applied to the investigation of DMA methodology. Our results have demonstrated that the performance of $m$th order centered DMA, where $m$ is a nonnegative even integer, is very well comparable with that of $(m+1)$th order DFA. It has been demonstrated that the zeroth order centered DMA has a good performance to characterize long-range correlation and fractal scaling behavior \cite{alvarez2005detrending, Shao2012}. In practical applications to real-world time series, the higher detrending power degree would be very important to improve estimation accuracy and to validate the observed scaling behavior \cite{kantelhardt2001detecting,bashan2008comparison}. Hence, our results would facilitate further application of higher order DMA.

\section*{Acknowledgements}

The author would like to thank Professors Taishin Nomura and Yasuyuki Suzuki for fruitful comments. This work was supported by JSPS KAKENHI Grant Number 15K01285 and 26461094.

\clearpage

\appendix

\section{Derivation of Eq.~(\ref{eq:SqDevt0})} \label{derivation_a0}

To obtain the least-squares polynomial by minimizing Eq.~(\ref{SqDevPoly1}), we
solve the following equations:
\begin{equation}
\frac{\partial I(\{a_i \})}{\partial a_j} = 0,
\end{equation}
where $i, j = 0,1,\cdots, m$. When $j$ is even ($j = 0,2,\cdots, m$), we get
\begin{equation}
\left[
\begin{array}{cccc}
\rho_0 & \rho_2 & \cdots & \rho_m \\
\rho_2 & \rho_4 & \cdots & \rho_{m+2} \\
\rho_4 & \rho_6 & \cdots & \rho_{m+4} \\
\vdots & \vdots & & \vdots \\
\rho_m & \rho_{m+2} & \cdots & \rho_{2m} \\
\end{array}
\right]\left[
\begin{array}{c}
a_0 \\
a_2 - \frac{c_1}{2} \\
a_4 - \frac{c_3}{4} \\
\vdots \\
a_m - \frac{c_{m-1}}{m} \\
\end{array}
\right]
=\left[
\begin{array}{c}
0 \\
0 \\
0 \\
\vdots \\
0
\end{array}
\right],
\end{equation}
where
\begin{equation}
\rho_k = 2 \int_{0}^{T/2} \!\! t^{k}\, dt = \frac{2^{-k}}{k+1} T^{k+1}.
\end{equation}
Because a unique solution should exist in the case of the least squares method, we obtain
\begin{eqnarray}
a_0 &=& 0 \\
a_{2k} &=& \frac{c_{2k-1}}{2k} \qquad {\rm for} \quad k = 1, 2, \cdots m/2.
\end{eqnarray}

\section{Exact calculation of the single-frequency response for discrete time series}\label{Appendix:EC}

To obtain the single-frequency response function, we have used a continuous time approximation. As we demonstrated in this paper, the single-frequency response function calculated under this assumption can help to gain deeper insight into the performance of DMA. However, it is also possible to calculate the exact form of the single-frequency response when discrete time series is analyzed, although the amount of calculation is somewhat large.
Here we provide some exact formulas of the single-frequency response function.

To obtain the single-frequency response function at scale $n=2n'+1$ for centered DMA, let us consider discrete-time series $\{x(i)\}$,
\begin{equation}
x(i) = A \cos \left(2 \pi f i + \theta \right),
\end{equation}
 in the range $[- n',n']$. In this case, the integrated series is given by
\begin{eqnarray}
y(i) &=& \sum_{j=-n'}^{i} x(i) \nonumber \\
 &=& \frac{A \sin \{ \pi  f (i+n+1) \} \sin \{\pi  f (i-n)+\theta \}}{\sin (\pi  f)}. \nonumber \\
\end{eqnarray}
To calculate the moving average polynomial at $i = 0$, we first obtain coefficients $\{a_k\}$ of the least-squares polynomial with degree $m$ by minimizing
\begin{equation}
I\left( \{a_j\} \right) = \sum_{i=-n'}^{n'} \left( y (i) - \sum_{k=0}^{m} a_k i^{k} \right)^2. \label{SqDevPoly}
\end{equation}
After the determination of $\{a_k\}$, the mean-square deviation at $i=0$ is given by
\begin{equation}
\Phi_{\rm d}^2 (n, f, A, \theta) =  \left( y (i) - \sum_{k=0}^{m} a_k i^{k} \right)^2.
\end{equation}
By averaging the phase $\theta$ in this equation over $[0, 2 \pi]$, we finally obtain the single-frequency response function $\overline{\Phi}_{\rm d}^2 (n, f, A)$ for the discrete-time series $\{x(i)\}$.

\begin{figure}[htbp]
       \begin{center}
               \includegraphics[width = 1\linewidth]{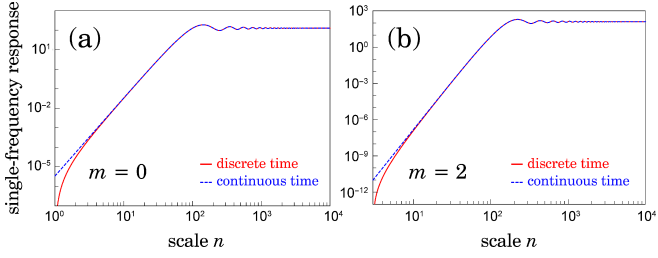}
               \caption{The single-frequency response
function of $m$th order centered DMA for $A=1$, and $f=0.01$. }
               \label{fig:exact_response}
       \end{center}
\end{figure}

For instance, in the zeroth order case, the exact form of the single-frequency response function is given by:
\begin{widetext}
\begin{eqnarray}
\overline{\Phi}_{\rm d}^2 (n, f, A) &=& \frac{A^2}{16 (2 n+1)^2 \sin^4(\pi f)} \Big[4 n^2 + 4n+2 - (2 n+1)^2 \cos (2 \pi  f)  \nonumber \\
&& +(4 n+2) \cos (2 \pi  f (n+1))-(4 n+2) \cos (2 \pi  f n)-\cos (2 \pi  f (2 n+1))\Big], \nonumber \\
\end{eqnarray}
and in the second order case, by
\begin{eqnarray}
\overline{\Phi}_{\rm d}^2 (n, f, A) &=& \frac{A^2}{128 (2 n+1)^2 \left(4 n^2+4 n-3\right)^2 \sin^8(\pi  f)} \Big[3 \left(8 n^3+12 n^2-2 n-3\right) \sin (\pi  f) \nonumber \\
&& +\left(-8 n^3-12 n^2+2 n+3\right) \sin (3 \pi f) + 3 \left(2 n^2+7 n+6\right) \sin (\pi  f (1-2 n)) \nonumber \\
&& +3 (2 n-1) ((2 n+3) \sin (\pi  f (2 n+1))-(n-1) \sin (\pi  f (2 n+3))) \Big]^2.
\end{eqnarray}

\end{widetext}
The functional forms of the single-frequency response functions for discrete-time and continuous-time cases are illustrated in Fig.~\ref{fig:exact_response}, where one can note that, except for differences at very small scales, the single frequency response function calculated under the continuous-time approximation in Section III is in excellent agreement with its exact result.

\section{Single-frequency response function of DMA}

\paragraph{Zeroth-order and first-order centered DMA} ($m=0, 1$)
\begin{eqnarray}
a_0 &=& \frac{A \sin \theta \, \sin (\pi  f T)}{2 \pi ^2 f^2 T}
\\
\overline{\Phi}^2 (T, f, A) &=& \frac{A^2 \left\{ \sin (\pi  f T)-\pi  f T \right\}^2}{8 \pi ^4 f^4 T^2} \\
&=& \frac{\pi ^2 A^2 f^2 T^4}{288}  + O \left(T^{6}\right) \quad (T \ll 1/f) \nonumber \\
\end{eqnarray}

\begin{widetext}
\paragraph{Second-order and Third-order centered DMA} ($m=2, 3$)
\begin{eqnarray}
a_0 &=& -\frac{3 A \sin \theta \, \left\{ \pi ^2 f^2 T^2 \sin (\pi  f T)-5 \sin (\pi  f T)+5 \pi  f T \cos (\pi  f T)\right\}}{4 \pi ^4 f^4 T^3}
\\
\overline{\Phi}^2 (T, f, A) &=& \frac{A^2 \left\{2 \pi ^3 f^3 T^3+3 \left(\pi ^2 f^2 T^2-5\right) \sin (\pi  f T)+15 \pi  f T \cos (\pi  f T)\right\}^2}{32 \pi ^8 f^8 T^6} \\
&=& \frac{\pi ^6 A^2 f^6 T^8}{627200} + O \left(T^{10}\right) \quad (T \ll 1/f) \end{eqnarray}

\paragraph{Fourth-order and Fifth-order centered DMA} ($m=4, 5$)
\begin{eqnarray}
a_0 &=& \frac{15 A \sin (\theta )}{16 \pi ^6 f^6 T^5} \left\{\pi ^4 f^4 T^4 \sin (\pi  f T)+14 \pi ^3 f^3 T^3 \cos (\pi  f T)-77 \pi ^2 f^2 T^2 \sin (\pi  f T)+189 \sin (\pi  f T) \right. \nonumber \\
&& \left. -189 \pi  f T \cos (\pi  f T) \right\} \nonumber  \\
\\
\overline{\Phi}^2 (T, f, A) &=& \frac{A^2}{512 \pi ^{12} f^{12} T^{10}} \left(-8 \pi ^5 f^5 T^5+105 \pi  f T \left(2 \pi ^2 f^2 T^2-27\right) \cos (\pi  f T) \right. \nonumber \\
&& \left. +15 \left(\pi ^4 f^4 T^4-77 \pi ^2 f^2 T^2+189\right) \sin (\pi  f T)\right)^2 \\
&=& \frac{\pi ^{10} A^2 f^{10} T^{12}}{8851949568} + O \left(T^{14}\right) \quad (T \ll 1/f) \end{eqnarray}

\end{widetext}


\end{document}